\documentclass[prl,twocolumn,aps,showpacs]{revtex4-1}
\usepackage{graphicx}
\usepackage{amsmath,amssymb,dsfont,bbold}

\begin{document}
\title{A Simple Derivation of the Fong-Wandzura Pulse Sequence}

\author{Daniel Zeuch and N.~E. Bonesteel}

\affiliation{Department of Physics and National High Magnetic
Field Laboratory, Florida State University, Tallahassee, FL 32310,
USA}

%\date{\today}

\begin{abstract}
We give an analytic construction of a class of two-qubit gate pulse sequences that act on five of the six spin-$\frac12$ particles used to encode a pair of exchange-only three-spin qubits.  Within this class, the problem of gate construction reduces to that of finding a smaller sequence that acts on four spins and is subject to a simple constraint. The optimal sequence satisfying this constraint yields a two-qubit gate sequence equivalent to that found numerically by Fong and Wandzura.  Our construction is sufficiently simple that it can be carried out entirely with pen, paper, and knowledge of a few basic facts about quantum spin. We thereby analytically derive the Fong-Wandzura sequence that has so far escaped intuitive explanation. 
\end{abstract}

%\pacs{03.67.Lx, 73.21.La}

\maketitle

Control of the Heisenberg exchange coupling between pairs of spin-$\frac12$ particles is a useful resource for carrying out quantum gates in a quantum computer \cite{loss98}. If qubits are encoded in the Hilbert space of one \cite{loss98} or two \cite{levy02} spin-$\frac12$ particles, resources beyond exchange are required for universal quantum computation.  However, controlled exchange alone is universal if each qubit is encoded in the Hilbert space of at least three spin-$\frac12$ particles \cite{bacon00,kempe01,divincenzo00}.  

Semiconductor quantum dots with trapped electrons are promising systems for manipulating spin-$\frac12$ particles \cite{hanson07}. Controlled exchange between pairs of electron spins in quantum dots has been demonstrated \cite{petta05}, and used to manipulate a variety of three-spin encoded qubits  \cite{laird10,gaudreau12,hsieh12,shi12,medford13_nn,eng15}, including the so-called resonant exchange qubit \cite{taylor13,medford13_prl,doherty13} which maintains qubit encoding by keeping the exchange interaction ``always on" within each qubit (see also \cite{weinstein05}). Here we focus on the case of exchange-only quantum computation where the exchange interaction is kept completely off except when being pulsed, i.e. adiabatically switched on and off, between pairs of spins.  It is then necessary to design pulse sequences that carry out quantum gates on encoded qubits without resulting in leakage out of the encoded qubit space \cite{divincenzo00,hsieh03,kawano05,fong11,shi12,setiawan14,zeuch14}.  

To assess any quantum computation scheme one ultimately needs to know the minimal cost of carrying out quantum gates.  For exchange-only quantum computation using pulse sequences, single-qubit gate sequences are theoretically understood but there is little true understanding regarding optimization of two-qubit gate sequences.  The main difficulty comes from the constraint of no leakage combined with the large search space of unitary operators acting on the six spins encoding a pair of three-spin qubits.  Not surprisingly, the shortest known pulse sequence for an entangling two-qubit gate due to Fong and Wandzura \cite{fong11} has been found by a numerical search algorithm which offers little insight into its derivation. Furthermore, existing analytic derivations of less optimal sequences are lengthy and complicated \cite{kawano05,zeuch14}.    

In this Letter we present an analytic construction of a class of pulse sequences that carry out two-qubit gates for exchange-only quantum computation.  These sequences are built out of smaller sequences that act on only four spins and satisfy a certain constraint.  We show that when the most efficient of these smaller sequences is used the result is equivalent to the Fong-Wandzura sequence.  Our guiding principle throughout is to avoid as much as possible complicated calculations and use only the most basic facts about quantum spin \cite{facts}.   

\begin{figure}
	\includegraphics[width=\columnwidth]{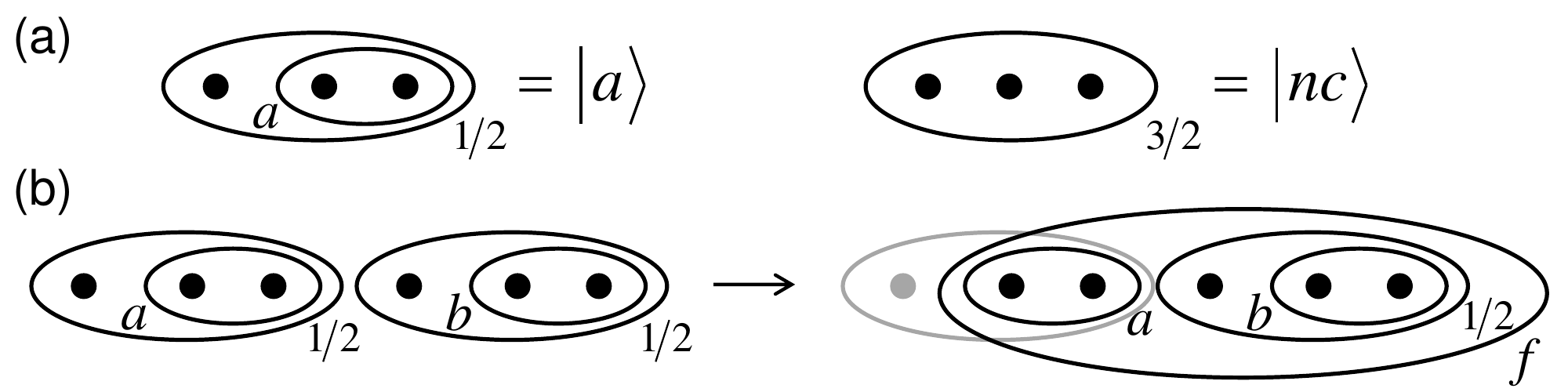}
	\caption{(a) Three-spin encoding for logical qubit states with total spin $\frac12$, $|a\rangle$ with $a=0,1$, and noncomputational state with total spin $\frac32$, $|nc\rangle$. (b) Two qubits in states $a$ and $b$ and a diagram highlighting the five rightmost spins with total spin $f=\frac12$ or $\frac32$.}
	\label{basics}
\end{figure}

Because we only consider the action of rotationally invariant operators, it is sufficient to describe quantum states of multiple spins using only total spin quantum numbers.  Accordingly, we employ the notation used in \cite{zeuch14} in which each spin-$\frac12$ particle is represented by the symbol $\bullet$ and groups of particles are enclosed in ovals labeled by the total spin of the enclosed particles. Figure \ref{basics}(a) shows the three-spin qubit encoding of \cite{divincenzo00} in this notation.  In the text we write these states using parentheses instead of ovals, so the computational qubit states shown in Fig.~\ref{basics}(a) are $(\bullet(\bullet\bullet)_a)_{1/2} = |a\rangle$ where $a=0$ and 1 define the standard basis. Figure \ref{basics}(a) also shows the non-computational state $(\bullet\bullet\bullet)_{3/2} = |nc\rangle$.  Figure \ref{basics}(b) shows six spins encoding two qubits and highlights the five spins the pulse sequences we consider act on, where the total spin $f$ can be either $\frac12$ or $\frac32$.   

\begin{figure}
	\includegraphics[width=\columnwidth]{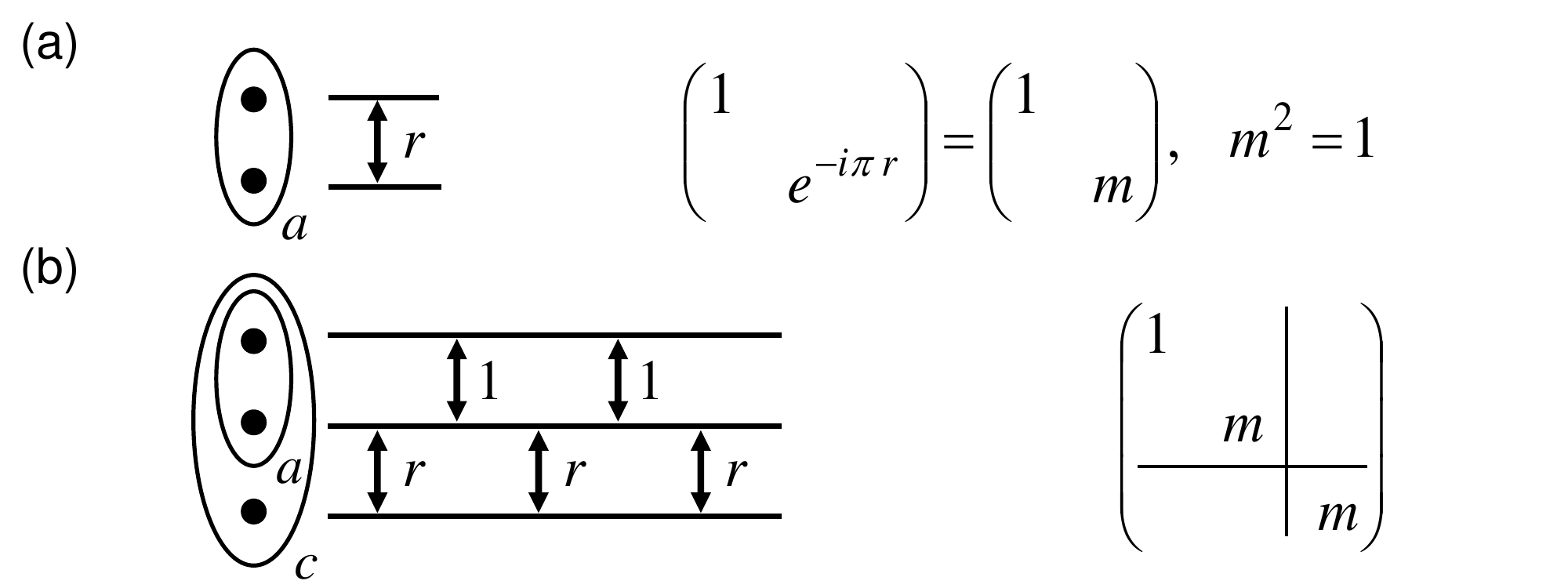}%
	\caption{(a) Elementary exchange pulse of duration $r=0$ or 1, referred to as an $r$-pulse in the text, and matrix representation of the resulting operation in the basis $a=\{0,1\}$, where $m=1$ or $-1$ for $r=0$ or 1, respectively.  (b) Sequence of three $r$-pulses and two explicit SWAP pulses ($r$-pulses with $r=1$) acting on three spins and the matrix representation of the resulting operation in the basis $ac=\{0\frac12,1\frac12|1\frac32\}$.}
	\label{observation1a}
\end{figure}

Consider the exchange Hamiltonian $H=J \mathbf S_i \cdot \mathbf S_j$ acting on two spin-$\frac12$ particles $(\bullet\bullet)_a$ whose Hilbert space is spanned by the two states with total spin $a=0$ and 1. (Because only total spin quantum numbers are relevant in our analysis, we treat, for example, the three-fold degenerate $a=1$ state as a single state.)  Pulsing this Hamiltonian for a duration $t$ measured in units of $1/(\pi J)$ results in the time evolution operator (up to an irrelevant overall phase factor), 
\begin{equation}
	U_{ij}(t) =\textnormal{diag}(1,e^{-i\pi t}),
	\label{exchange}
\end{equation}
where the matrix representation is given in the $a=\{0,1\}$ basis.  We take $t \in [0,2)$ for which the inverse pulse has duration $2-t$. 

Two exchange pulses square to the identity and play a fundamental role in our construction.  The durations of these pulses, which we denote $r$, can only be either 0 or 1, and we refer to them as $r$-pulses.  For $r=0$ an $r$-pulse is simply the identity, while for $r=1$ it is a SWAP operation, which is equivalent to physically exchanging two spins \cite{phase}.  Figure \ref{observation1a}(a) shows an $r$-pulse acting on two spins in the state $(\bullet\bullet)_a$ using standard notation with exchange pulses represented by double arrows labeled by duration.  The corresponding matrix representation of the resulting operation, also given in Fig.~\ref{observation1a}(a), shows that applying an $r$-pulse multiplies the $a=0$ state by 1 and the $a=1$ state by $m$, where $m=1$ or $m=-1$ for $r=0$ or $r=1$, respectively. In both cases $m^2 = 1$, reflecting the fact that the $r$-pulses square to the identity.

A pulse sequence that acts on three spins and consists of three $r$-pulses (with either $r=0$ or 1) acting on the bottom two spins and two explicit SWAP pulses acting on the top two spins is shown in Fig.~\ref{observation1a}(b).  For $r=0$ the two explicit SWAP pulses square to the identity. For $r=1$ the sequence consists of five SWAP pulses which, when viewed as spin permutations, are readily seen to be equivalent to a single SWAP pulse acting on the top two spins.  In both cases, the effect of the sequence is to multiply the state $((\bullet\bullet)_a\bullet)_c$ by 1 if $a=0$, and $m$ if $a=1$, regardless of the value of $c$, where, as in Fig.~\ref{observation1a}(a), $m=1$ or $m=-1$ for $r=0$ or $r=1$, respectively.  The corresponding matrix representation is also given in Fig.~\ref{observation1a}(b). 

\begin{figure}
	\includegraphics[width=\columnwidth]{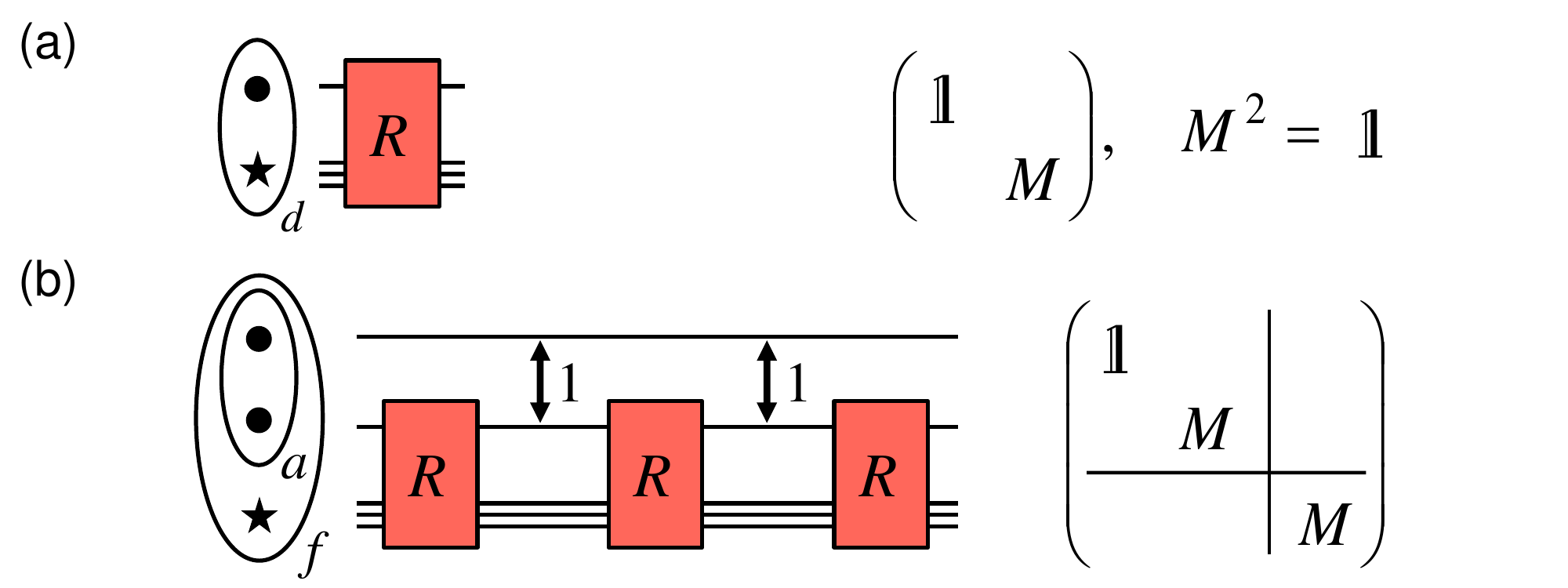}
	\caption{(Color online) (a) Operation, $R$, acting on a spin-$\frac12$ particle, $\bullet$, and an effective spin-$\frac12$ particle, $\bigstar$, which represents the three spins encoding the target qubit. The matrix representation of this operation, which generalizes the $r$-pulse of Fig.~\ref{observation1a}(a), is given in the effective basis $d=\{0,1\}$ where $\mathbb{1}$ is the $2\times 2$ identity, and $M$ is a $2\times 2$ matrix where $M^2 = \mathbb{1}$, both of which act on the target qubit contained in $\bigstar$.  (b) Pulse sequence which generalizes that shown in Fig.~\ref{observation1a}(b).  The matrix representation of the resulting operation is given in the effective basis $af=\{0\frac12,1\frac12|1\frac32\}$.}
	\label{observation1b}
\end{figure}

We seek pulse sequences which act on the five spins $((\bullet\bullet)_a(\bullet(\bullet\bullet)_b)_{1/2})_f$ highlighted in Fig.~\ref{basics}(b) and carry out leakage-free two-qubit gates.  For reasons that will become clear we refer to the qubit with state label $a$ as the control qubit and the qubit with state label $b$ as the target qubit. Our construction is based on using a smaller sequence which acts on the four rightmost spins in Fig.~\ref{basics}(b).  This smaller sequence carries out an operation we denote $R$ which, as seen shortly, is closely related to an $r$-pulse.  One requirement we place on $R$ is that it not result in any leakage of the target qubit into its noncomputational state. We can therefore work within an effective Hilbert space in which the three spins encoding the target qubit are replaced by a single effective spin-$\frac12$ particle,
\begin{equation}
	(\bullet(\bullet\bullet)_b)_{1/2} \rightarrow \bigstar.
	\label{bigstar}
\end{equation}
Matrix elements of operations acting on any collection of spins including $\bigstar$ are then elevated from numbers to $2\times 2$ blocks that act on the Hilbert space of the target qubit hidden within $\bigstar$. 

We require that when $R$ is applied to the state $(\bullet\bigstar)_d$ it act on the target qubit with the identity $\mathbb{1}$ if $d=0$, and a matrix $M$, with $M^2 = \mathbb{1}$, if $d=1$, as also shown in the corresponding matrix representation of $R$ given in Fig.~\ref{observation1b}(a).  Such $R$ operations can be viewed as generalized $r$-pulses where the matrix elements 1 and $m$, with $m^2 = 1$, of Fig.~\ref{observation1a}(a) have been elevated to the $2\times 2$ matrices $\mathbb{1}$ and $M$, with $M^2= \mathbb{1}$ in Fig.~\ref{observation1b}(a).

This view of $R$ as an elevated $r$-pulse suggests the five-pulse sequence of Fig.~\ref{observation1a}(b) can also be elevated to the sequence shown in Fig.~\ref{observation1b}(b). This sequence acts on the effective Hilbert space spanned by the states $((\bullet\bullet)_a\bigstar)_f$ with $af=0\frac12$, $1\frac12$ and $1\frac32$ and consists of three $R$ operations acting on the central spin and the effective spin $\bigstar$ and two SWAP pulses acting on the top two spins.  The only $2\times 2$ block element in the matrix representation of $R$ which is not proportional to the identity is $M$.  Because $M^2 = \mathbb{1}$, when evaluating the matrix representation for the full sequence, each block matrix element must be of the form $\alpha_0\ \mathbb{1}+\alpha_1\ M$.  The coefficients $\alpha_0$ and $\alpha_1$ for each block element are completely determined by the two cases $M=\pm \mathbb{1}$, which are equivalent to the cases $m=\pm1$ in Fig.~\ref{observation1a}(b).  It follows that the matrix representation of the operation carried out by this sequence in the effective $af = \{0\frac12, 1\frac12|1\frac32\}$ basis is that given in Fig.~\ref{observation1b}(b), i.e. an elevated version of the matrix shown in Fig.~\ref{observation1a}(b). To prove this result we have only used the fact that $M^2 = \mathbb{1}$.  It therefore holds not just for the straightforward cases $M = \pm \mathbb{1}$, but also for any matrix of the form $M = \mathbf {\hat n} \cdot \boldsymbol\sigma$ where $\mathbf{\hat n}$ is a real-valued unit vector and $\boldsymbol{\sigma}=(\sigma_x,\sigma_y,\sigma_z)$ is the Pauli vector. 

The pulse sequence shown in Fig.~\ref{observation1b}(b) acting on the two qubits of Fig.~\ref{basics}(b) applies the identity $\mathbb{1}$ to the target qubit when the state of the control qubit is $a=0$, and applies the matrix $M$ to the target qubit when the state of the control qubit is $a=1$, regardless of the value of $f$.  The matrix representation of the operation carried out by this sequence can then be given in the standard two-qubit basis $ab=\{00,01,10,11\}$ as
\begin{equation}
	U_{2qubit}=\textnormal{diag}(\mathbb{1},M).
	\label{two-qubit}
\end{equation} 
For $M= \pm \mathbb{1}$ the resulting gate is not entangling. However, for $M=\mathbf{\hat n}\cdot\boldsymbol\sigma$ the sequence enacts a leakage-free controlled-$(\mathbf{\hat n}\cdot \boldsymbol\sigma)$ gate which is equivalent to a controlled-NOT (CNOT) gate (for which $\mathbf{\hat n} = \mathbf{\hat x}$), up to single-qubit rotations.

Abandoning the notation $\bigstar$ we now consider $R$ acting on the four-spin Hilbert space spanned by the states $(\bullet((\bullet\bullet)_b\bullet)_c)_d$ where, since $c$ is initially $\frac12$, $d$ can only be either 0 or 1.  The requirements on $R$ needed to construct a controlled-$(\mathbf{\hat n}\cdot \boldsymbol\sigma)$ gate are that it must i) preserve the quantum number $c$, and ii) in the restricted Hilbert space with $c=\frac12$, have the form shown in Fig.~\ref{observation1b}(a) with $M = \mathbf{\hat n}\cdot \boldsymbol\sigma$.

To construct a sequence for $R$ we introduce a new operation $V$ which satisfies the constraint
\begin{eqnarray}
	\langle ((\bullet\bullet)_1(\bullet\bullet)_{1})_1 | V |(\bullet(\bullet\bullet\bullet)_{3/2})_1\rangle = 0 \label{constraint}
\end{eqnarray}
depicted in Fig.~\ref{observation2}(a).  As shown below, inserting any $V$ satisfying (\ref{constraint}) into the sequence shown in Fig.~\ref{observation2}(b) results in an $R$ operation with $M = \mathbf{\hat n}\cdot \boldsymbol\sigma$. Letting $U_{ij}(t)$ denote an exchange pulse of duration $t$ acting on spins $i$ and $j$, as defined in (\ref{exchange}),  the sequence for $R$ can be written $V^{-1} U_{12}(1) U_{34}(1) V$, using the spin labeling of Fig.~\ref{observation2}(b). The matrix representation of the central two SWAP pulses $U_{12}(1) U_{34}(1)$ in the $((\bullet\bullet)_{b^\prime}(\bullet\bullet)_b)_d$ basis with state ordering $b b^\prime d = \{000,110|011,101,111\}$ is,
\begin{eqnarray}
U_{12}(1) U_{34}(1) = \textnormal{diag}(1,1|-1,-1,1).\label{swaps}
\end{eqnarray}
In the $d=0$ sector $U_{12}(1) U_{34}(1)$ acts as the identity, and thus $R$ also acts as the identity since $V$ and $V^{-1}$ cancel one another.  In the $d=1$ sector, (\ref{constraint}) implies that $V$ maps the $c=\frac32$ state $(\bullet (\bullet\bullet\bullet)_{3/2})_1$ entirely into the $b^\prime b = 01,10$ subspace.  The two central SWAP pulses then apply a phase factor of $-1$ to any state in this subspace, and so after applying $V^{-1}$ the net effect of the full sequence will be to multiply the $c=\frac32$ state by $-1$.  The fact that $R$ maps the $c=\frac32$ state onto itself immediately implies that $R$ also maps the $c=\frac12$ subspace onto itself, and thus leads to no leakage of the target qubit.  Furthermore, since the trace of $U_{12}(1)U_{34}(1)$ in the $d=1$ sector is $-1$ (see (\ref{swaps})), the trace of the full sequence $V U_{12}(1) U_{34}(1) V^{-1}$ in this sector is also $-1$. Thus, since the $c=\frac32$ matrix element of the full sequence is $-1$, the trace of the operation acting on the $c=\frac12$ subspace must be 0.  Finally, because the two SWAP pulses $U_{12}(1) U_{34}(1)$ square to the identity the full sequence for $R$ squares to the identity.  The operation carried out by this sequence on the $c=\frac12$ subspace in the $d=1$ sector must therefore also square to the identity and, because it is traceless, it must have the form $M = \mathbf{\hat n}\cdot \boldsymbol\sigma$.

\begin{figure}
	\includegraphics[width=\columnwidth]{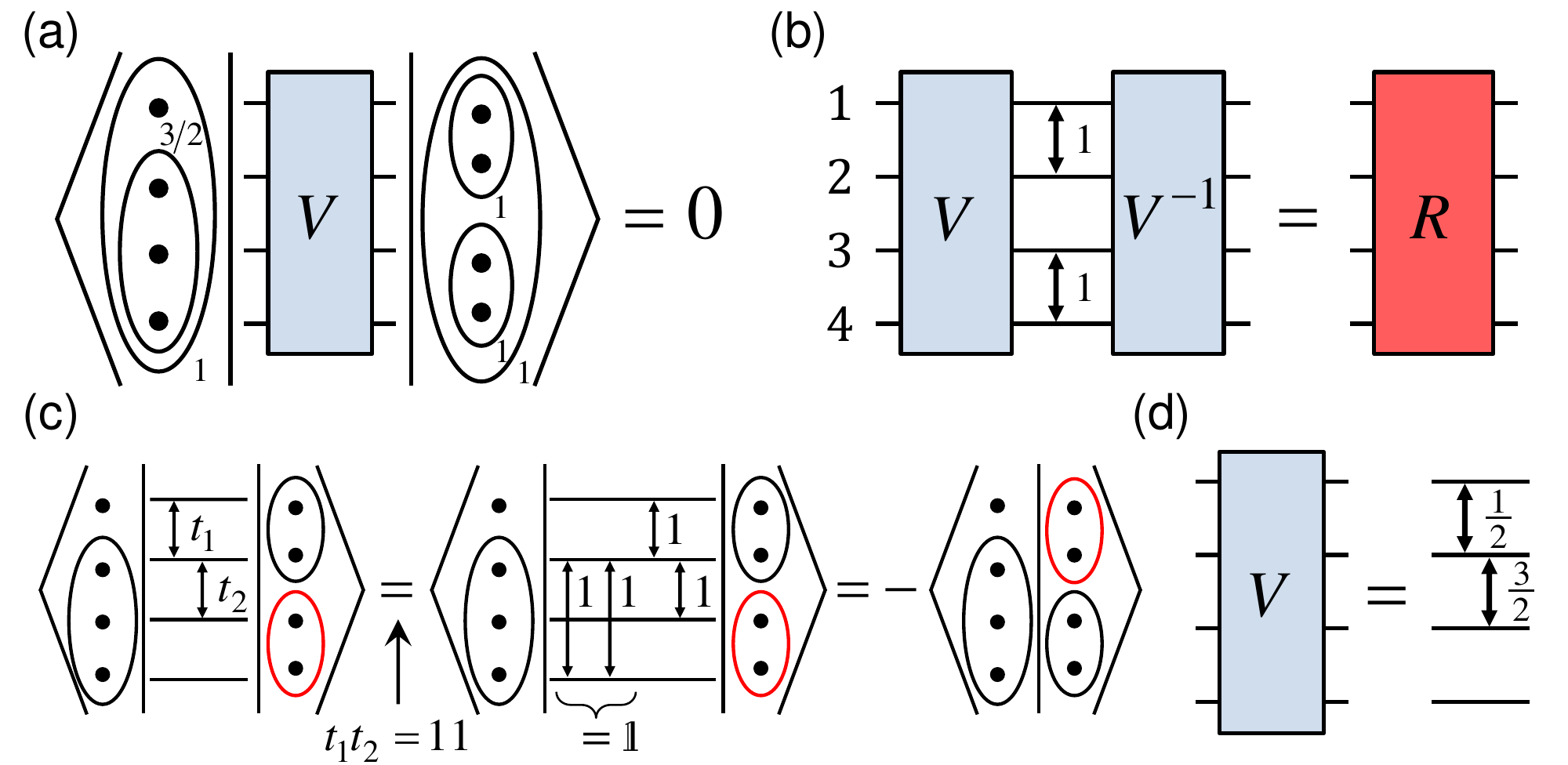}
	\caption{(Color online) Constructing a pulse sequence for $R$.  (a) Constraint which must be satisfied by an operation, $V$, used in the construction.  (b) Sequence in which $V$, its inverse, and two SWAP pulses are used to carry out an $R$ operation. (c) Evaluation of the matrix element (\ref{overlap}) for the case $t_1t_2 = 11$, as described in the text. (All quantum numbers are the same as those in (a), but are omitted for readability.) (d) One of two two-pulse solutions of (a) for $V$ which can be used to construct $R$ using (b).}
	\label{observation2}
\end{figure}

The set of pulse sequences $V$ that satisfy the constraint (\ref{constraint}) can be used to construct an infinite class of sequences resulting in two-qubit gates which are locally equivalent to CNOT.  We now show that the fewest number of pulses needed to satisfy (\ref{constraint}) is two and for this optimal case the resulting two-qubit gate sequence is equivalent to the Fong-Wandzura sequence. 

Without loss of generality we take $V = U_{23}(t_2) U_{12}(t_1)$ \cite{wlg}.  While the values of $t_1$ and $t_2$ for which $V$ satisfies (\ref{constraint}) can be found by brute force calculation, we determine them here by a simple ``pen and paper" procedure.  This procedure is based on the observation that (\ref{exchange}) implies 
\begin{eqnarray}
	\langle ((\bullet\bullet)_1(\bullet\bullet)_{1})_1 | U_{23}(t_2)U_{12}(t_1) |(\bullet(\bullet\bullet\bullet)_{3/2})_1\rangle = \hspace{1cm}\nonumber\\
	\alpha+\beta e^{-i\pi t_1}+\gamma e^{-i\pi t_2}+\delta e^{-i\pi (t_1+t_2)},\hspace{.6cm}
	\label{overlap}
\end{eqnarray}
where the coefficients $\alpha$, $\beta$, $\gamma$, and $\delta$ can be found by evaluating the four simple cases $t_1t_2 = 00, 01, 10,$ and 11. For the case $t_1t_2 = 00$ the matrix element (\ref{overlap}) is simply equal to the overlap $F \equiv \langle ((\bullet\bullet)_1(\bullet\bullet)_{1})_1 | (\bullet(\bullet\bullet\bullet)_{3/2})_1\rangle$ \cite{F}. For both cases $t_1t_2=01$ and 10 there is a single SWAP pulse which can be applied either directly to the left (for $t_1t_2 = 01$) or right (for $t_1t_2=10$) four-spin state in Fig.~\ref{observation2}(a).  Since each SWAP pulse then acts on a pair of spins with total spin 1, the result is an overall $-1$ using the phase convention of (\ref{exchange}).  In both cases the matrix element (\ref{overlap}) is thus equal to $-F$.  

The only non-trivial case is $t_1t_2=11$ for which both pulses are SWAP pulses.  A method for evaluating (\ref{overlap}) in this case is sketched in Fig.~\ref{observation2}(c).  First, a pair of SWAP pulses which combine to the identity, $U_{24}(1) U_{24}(1) = \mathbb{1}$, is inserted at the start of the sequence.  We then view the four SWAP pulses as physical particle exchanges.  It is irrelevant that the effect of particle exchange differs from that of a SWAP pulse by a factor of $-1$ because there are an even number of SWAP pulses.  Applying one of the exchanges acting on spins 2 and 4 to the state $(\bullet(\bullet\bullet\bullet)_{3/2})_1$ then gives a factor $+1$, since the two spins being exchanged have total spin 1.  The remaining three exchanges can then be applied to the state $((\bullet\bullet)_1(\bullet\bullet)_1)_1$ where, referring to the Fig.~\ref{observation2}(c), they result in a permutation which exchanges the bottom two spins (red oval) with the top two spins (black oval).  The net effect is therefore to exchange two spin-1 objects with total spin 1 and, as a result, the right state acquires a factor of $-1$.  Thus the $t_1t_2=11$ matrix element (\ref{overlap}) is equal to $-F$.  

Having evaluated the left-hand side of (\ref{overlap}) for the four cases $t_1t_2 = 00$,01,10, and 11, the coefficients appearing on the right-hand side are easily found to be $-\alpha=\beta=\gamma=\delta=F/2$.  For these coefficients there are only two solutions which satisfy (\ref{constraint}), $t_1t_2=\frac12\frac32, \frac32\frac12$.  Figure \ref{observation2}(d) shows the resulting sequence for the first solution, which consists of one $\sqrt{\text{SWAP}}$ $(t=\frac12)$ and one inverse $\sqrt{\text{SWAP}}$ $(t=\frac32)$ pulse.  

\begin{figure}[t]
	\includegraphics[width=\columnwidth]{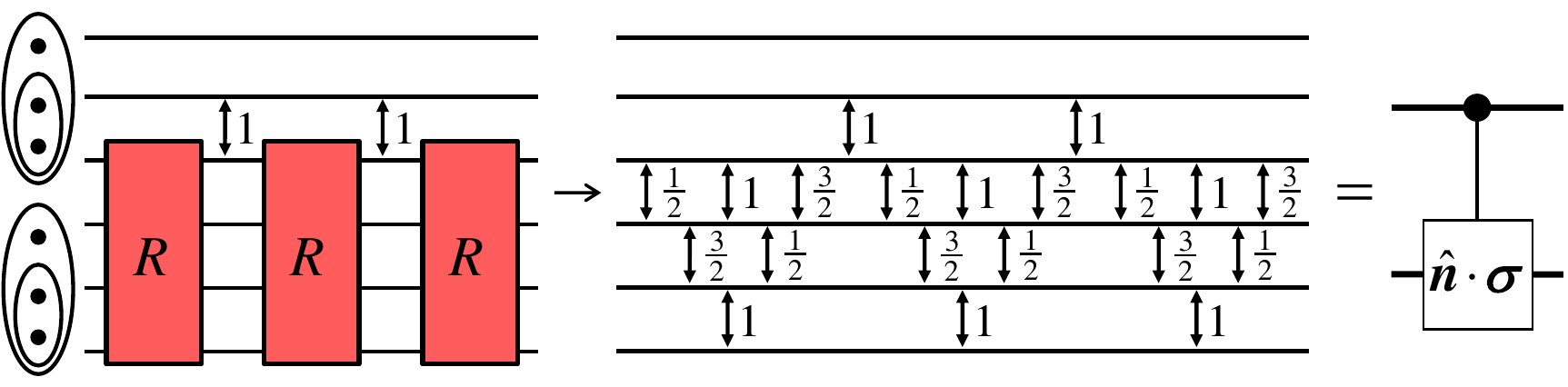}
	\caption{(Color online) Sequence of Fig.~\ref{observation1b}(b) acting on two encoded qubits resulting in a leakage-free controlled-$(\mathbf{\hat n}\cdot \boldsymbol\sigma)$ gate.  Also shown is the full sequence obtained by first inserting the optimal sequence for $V$ from Fig.~\ref{observation2}(d) into the sequence for $R$ from Fig.~\ref{observation2}(b) and inserting the result into the sequence from Fig.~\ref{observation1b}(b).  This sequence is equivalent to the Fong-Wandzura sequence.}
	\label{sequence}
\end{figure}

Figure \ref{sequence} shows the pulse sequence obtained by inserting the sequence for $V$ from Fig.~\ref{observation2}(d) into Fig.~\ref{observation2}(b) and inserting the resulting sequence for $R$ into Fig.~\ref{observation1b}(b). This sequence can be applied to a linear array of spins with nearest-neighbor pulses and carries out a controlled-$(\mathbf{\hat n}\cdot \boldsymbol\sigma)$ gate consisting of eight SWAP pulses, six $\sqrt{\text{SWAP}}$ pulses, and six inverse $\sqrt{\text{SWAP}}$ pulses \cite{n}. After single-qubit rotations are removed, the core Fong-Wandzura sequence as published in \cite{fong11} consists of six SWAP pulses, three $\sqrt{\text{SWAP}}$ pulses, and nine inverse $\sqrt{\text{SWAP}}$ pulses.  Our sequence can be converted into the Fong-Wandzura sequence through a series of elementary manipulations in which pairs of SWAP pulses are inserted (as in Fig.~\ref{observation2}(c)) or removed from the sequence, and single SWAP pulses are pulled past other pulses and in some cases combined with $\sqrt{\text{SWAP}}$  pulses to form inverse $\sqrt{\text{SWAP}}$ pulses (and vice versa).  The same manipulations can be used to produce sequences applicable to spin geometries other than linear \cite{setiawan14}.  These manipulations preserve the fact that there are twelve nontrivial (i.e. not SWAP) pulses in these sequences as well as the parity of the sum of the number of SWAP and $\sqrt{\text{SWAP}}$ pulses.  This parity is odd for the Fong-Wandzura sequence and even for our construction, consistent with the fact that a single-qubit operation corresponding to a single SWAP pulse must be added to the core Fong-Wandzura sequence to produce a controlled-$(\mathbf{\hat n}\cdot \boldsymbol\sigma)$ gate.  

In summary, we have analytically constructed a class of pulse sequences for carrying out two-qubit gates for exchange-only quantum computation.  These sequences can be viewed as elevated versions of the simple three-spin sequences shown in Fig.~\ref{observation1a}(b) which consist entirely of SWAP operations.  To carry out this elevation we introduced the four-spin sequence $R$ which is itself built out of a smaller sequence $V$ which satisfies the constraint (\ref{constraint}).  When the shortest pulse sequence for $V$ is plugged back into the full two-qubit sequence the result is equivalent to the Fong-Wandzura sequence. 

This work was supported in part by the U.S. DOE Grant No. DE-FG02-97ER45639.

\bibliography{bibliography}

\begin{thebibliography}{27}
\bibitem [{fac()}]{facts}%
  \BibitemOpen
  \href@noop {} {}\bibinfo {note} {The essential ingredients in our
  construction are the familiar spin addition rule, $s_1 \otimes s_2 = |s_1 -
  s_2|, |s_1 - s_2| + 1 \cdots, s_1+s_2$, and the fact that interchanging two
  spin-$s$ particles with total spin $S_{tot}$ results in a phase factor of
  $(-1)^{2s - S_{tot}}$.}\BibitemShut {Stop}%
\bibitem [{pha()}]{phase}%
  \BibitemOpen
  \href@noop {} {}\bibinfo {note} {For the phase choice in (\ref{exchange}) a
  SWAP pulse is equivalent to carrying out a physical particle exchange and
  multiplying the resulting state by $-1$.}\BibitemShut {Stop}%
\bibitem [{wlg()}]{wlg}%
  \BibitemOpen
  \href@noop {} {}\bibinfo {note} {Any $V = U_{kl}(t_2) U_{ij}(t_1)$ for which
  $[U_{ij}(t_1),U_{kl}(t_2)] \neq 0$,
  $(\bullet(\bullet\bullet\bullet)_{3/2})_1$ is not an eigenstate of
  $U_{ij}(t_1)$, and $((\bullet\bullet)_1(\bullet\bullet)_1)_1$ is not an
  eigenstates of $U_{kl}(t_2)$ will result in a two-qubit sequence equivalent
  to that obtained for the choice made in the text.}\BibitemShut {Stop}%
\bibitem [{F()}]{F}%
  \BibitemOpen
  \href@noop {} {}\bibinfo {note} {The value of $F$ can be found by direct
  calculation to be $1/\sqrt{3}$, up to a phase factor that is equal to $1$ for
  the Condon-Shortley convention.}\BibitemShut {Stop}%
\end{thebibliography}

\end{document}